\documentclass[twocolumn,nofootinbib,amsmath,amssymb,aps,prd,balancelastpage,superscriptaddress]{revtex4-1}

\usepackage{color}
\usepackage{xcolor}
\usepackage[active]{srcltx}
\usepackage{amsmath,amsfonts,amssymb,amsthm,amstext,amscd,eucal,srcltx}
\usepackage{epsfig,graphicx,bm}
\usepackage{epstopdf, epsf}
\usepackage{dcolumn}
\usepackage{comment}
\usepackage{hyperref}

\newcommand{\be}{\begin{equation}}
\newcommand{\ee}{\end{equation}}

\newcommand{\bse}{\begin{subequations}}
\newcommand{\ese}{\end{subequations}}
\newcommand{\bea}{\begin{eqnarray}}
\newcommand{\eea}{\end{eqnarray}}
\newcommand{\ba}{\begin{array}}
\newcommand{\ea}{\end{array}}
\newcommand{\bc}{\begin{center}}
\newcommand{\ec}{\end{center}}

\begin{document}
\preprint{IPM/P-2012/009}  
\vspace*{3mm}

\title{Experimental test of Non-Commutative Quantum Gravity by VIP-2 Lead}%

\author{{Kristian Piscicchia$^{b,c}$, Andrea Addazi$^{a,c \dagger}$, Antonino Marcian\`{o}$^{d,c \star}$, Massimiliano Bazzi$^c$, Michael Cargnelli$^{e,c}$,  Alberto Clozza$^c$, Luca De Paolis$^c$, Raffaele Del Grande$^{f,c}$, Carlo Guaraldo$^c$, Mihail Antoniu Iliescu$^c$, Matthias Laubenstein$^g$, Johann Marton$^{e,c}$, Marco Miliucci$^c$, Fabrizio Napolitano$^c$, Alessio Porcelli$^{e,c}$, Alessandro Scordo$^c$, Diana Laura Sirghi$^{c,h}$, Florin Sirghi$^{c,h}$, Oton Vazquez Doce$^c$, Johann Zmeskal$^{e,c}$ and Catalina Curceanu$^{c}$}\\ 
\vspace{0.5 cm}
{\it$^a$ Center for Theoretical Physics, College of Physics Science and Technology, Sichuan University, 610065 Chengdu, China}\\
{\it$^\dagger$addazi@scu.edu.cn}\\
{\it$^b$ Centro Ricerche Enrico Fermi - Museo Storico della Fisica e Centro Studi e Ricerche “Enrico Fermi”, Roma, Italy, EU}\\
{\it$^c$ Laboratori Nazionali di Frascati INFN, Frascati (Rome), Italy, EU}\\
{\it$^d$ Center for Field Theory and Particle Physics \& Department of Physics\\
Fudan University, Shanghai, China} \\
{\it$^\star$marciano@fudan.edu.cn} \\
{\it$^e$ Stefan Meyer Institute for subatomic physics, Austrian Academy of Science, Austria, EU}\\
{\it$^f$ Physik Department E62, Technische Universität München, 85748 Garching, Germany, EU}\\
{\it$^g$ Laboratori Nazionali del Gran Sasso INFN, Assergi (L'Aquila), Italy, EU}\\
{\it$^h$ IFIN-HH, Institutul National pentru Fizica si Inginerie Nucleara Horia Hulubei, Romania, EU}
}

\begin{abstract}
\noindent
Pauli Exclusion Principle (PEP) violations induced by space-time non-commutativity, a class of universality for several models of Quantum Gravity, are investigated by the VIP-2 Lead experiment at the Gran Sasso underground National Laboratory of INFN. The VIP-2 Lead experimental bound on the non-commutative space-time scale $\Lambda$ excludes $\theta$-Poincar\'e far above the Planck scale for non vanishing ``electric-like" components of $\theta_{\mu \nu}$, and up to $6.9 \cdot 10^{-2}$ Planck scales if they are null. Therefore, this new bound represents the tightest one so far provided by atomic transitions tests.
This result strongly motivates high sensitivity underground X-ray measurements as critical tests of Quantum Gravity and of the very microscopic space-time structure. 
\end{abstract}

\maketitle

\section{Introduction}\label{intro}
\noindent 
The Pauli Exclusion Principle (PEP) is one of the main pillars of quantum mechanics, so deeply rooted in the foundations of physics to be the main responsible for the stability of atoms, nuclei, molecules and matter in general. PEP forbids fermions to occupy the same quantum mechanical state, avoiding an arbitrary number of electrons or nucleons on the same orbital. 
It is worth reminding that PEP is a direct consequence of the Spin-Statistics theorem (SST), as  proved by the same {\it W. Pauli} \cite{Pauli:1940zz}: it arises from anti-commutation rules of fermionic spinor fields, in the construction of the Fock space of the theory. 

The SST is based on Lorentz invariance as a fundamental assumption. This means that the PEP is directly related to the fate of the space-time symmetry and structure. Nonetheless, Lorentz Symmetry may be dynamically broken at a very high energy scale, without this phenomenon translating into a fundamental breakdown of the symmetry. In this case the generation of non-renormalizable operators, suppressed as inverse powers of the Lorentz violation scale $\Lambda$, 
is expected.
On the other hand approaches to Quantum Gravity, for which space-time coordinates {\it do non commute} close to the Planck scale (about $10^{19}\, {\rm GeV}$), thus deforming the Lorentz algebra at the very fundamental level, were put forward. The idea that the space-time might be non-commutative is usually accredited to {\it W. Heisenberg} (see e.g. Ref.~\cite{Heisenberg}), as an extension of the uncertainty principle, having been later elaborated by {\it H. Snyder} and {\it C.N. Yang} in Refs.~\cite{Snyder,Yang}. From a symplectic-geometry approach \cite{SymGeo} it is possible to unveil the deep relation intertwining space-time symmetries, statistics  \cite{Arzano:2007gr} and the uncertainty principle \cite{SymGUP}, hence providing concrete path-ways for falsification.
 
Non-commutativity of space-time is common to several Quantum Gravity frameworks, to which we refer as Non-Commutative Quantum Gravity models (NCQG). The connection of space-time non-commutativity with both String Theory (ST) \cite{Frohlich:1993es,Chamseddine1,Frohlich:1995mr,Connes:1997cr,Seiberg:1999vs} and Loop Quantum Gravity (LQG) \cite{AmelinoCamelia:2003xp,Freidel:2005me,Cianfrani:2016ogm,Amelino-Camelia:2016gfx, NClimLQG,BRAM, Brahma:2017yza} was extensively studied in literature\footnote{The fact that non-commutativity emerges in both theories may not be a coincidence, but can be conjectured to be instead related to a newly formulated ${\bf H}$-duality --- see e.g. Refs.~\cite{Addazi:2017qwt,Addazi:2018cyn} --- since a self-dual LQG formulation can be obtained from topological M-theory.
Besides ST and LQG contexts, deformed symmetries may effectively emerge from several other non-perturbative models of quantum geometry.}. 

The two main classes of non-commutative space-time models embedding deformed Poincar\'e symmetries are characterized by $\kappa$-Poincar\'e \cite{Agostini:2006nc,13, AmelinoCamelia:2007uy, Arzano:2016egk} and $\theta$-Poincar\'e \cite{AmelinoCamelia:2007wk,AmelinoCamelia:2007rn,Addazi:2017bbg,Addazi:2018jmt,Addazi:2019ruk,Addazi:2018ioz} symmetries. 
Among these latter, there exists a sub-class of models which preserves
unitarity of the S-matrix in the Standard Model sector  
\cite{AG,Addazi:2018jmt}. 
From the experimental point of view, the most intriguing prediction of this class of non-commutative models is an energy dependent probability ($\delta^2(E)$) for electrons to perform PEP violating atomic transitions. For both $\kappa$ and $\theta$ Poincar\'e the PEP violation probability turns to be of order one in the deformation parameter, when the probed energy is close to the scale of non-commutativity $\Lambda$ (nonetheless, trivial dimensional arguments are sufficient to show that first order results in $\theta$ Poincar\'e are suppressed by the square of the energy scale).
In the low energy regime, i.e. for energies much smaller than the non-commutativity scale, the PEP violation probability is highly suppressed, accounting for the lack of evidence of PEP violation signals over decades of experimental efforts in this direction.

The experimental tests of the Spin-Statistics connection are based on the search for signals of PEP violating processes, by exploiting different techniques, which may or not respect the Messiah and Greenberg (MG) \cite{messiah} superselection rule. According to MG, even assuming small mixed symmetry components in a primarily antisymmetric wave function, in a system composed by a fixed number of particles the symmetry of the world Hamiltonian would prevent transitions among two different symmetry states, i.e. electrons or nucleons would not perform PEP forbidden transitions to lower orbits.

PEP tests for electrons, respecting the constraint imposed by MG, are performed by introducing new electrons in a pre-existing system of electrons and constraining the probability that the newly formed state is symmetric. This was accomplished using various methods:

\begin{itemize}
    \item the estimate of the primordial abundance of anomalous $^5$Li, with three protons in the lowest level, leads to the limit ($\delta^2 < 2 \cdot 10^{-28}$)~\cite{thoma1992}, which is also the strongest constraint inferred from astrophysical and cosmological arguments; 

\item capture of $^{14}$C $\beta$ rays onto Pb atoms ($\delta^2 < 3 \cdot 10^{-2}$) \cite{Goldhaber1948}; 

\item pair production electrons captured on Ge ($\delta^2 < 1.4 \cdot 10^{-3}$) \cite{Elliott:2011cx}; 

\item search for PEP violating atomic transitions in conducting targets,
a prototype experiment of this class has been
carried out in Ref \cite{ramberg1990} following a
suggestion of Greenberg and Mohapatra \cite{Greenberg:1987aa}, it consists in measuring the X-ray emission from a target strip where a direct current is circulated
and looking for a difference in the spectra acquired with current on and off (best upper limit $\delta^2 < 8.6 \cdot 10^{-31}$) \cite{curceanu2017,napolitano2022}; 

\item a generalized version of the latter experimental technique consists in using as test fermions the
free electrons residing in the conduction band of the target, an improved version of the original analysis proposed in Ref. \cite{Elliott:2011cx}, based on the calculation of the time dependence of the anomalous
X-ray emission process, lead to the best upper limit $\delta^2 < 1.53 \cdot 10^{-43}$ \cite{Piscicchia:2020kze}.
\end{itemize}

NCQG models are not subject to the MG superselection rule. Within this context the prototype Reines-Sobel experiment ~\cite{reines} looked for anomalous transitions in a stable system, i.e. spontaneous PEP-violating transition in a closed system of electrons in an established symmetry state. In this framework  strong bounds on the PEP violation probability were set by the DAMA/LIBRA collaboration ($\delta^2 < 1.28 \cdot 10^{-47}$) \cite{bernabei2009}, searching for K-shell PEP violating transitions in iodine 
--- see also Refs.~\cite{ejiri}. A similar strategy was followed by the MALBEK experiment \cite{abgrall}, which set the bound $\delta^2 < 2.92 \cdot 10^{-47}$ by constraining the rate of K$_\alpha$ PEP violating transitions in Germanium. 

This research is based on a different strategy, which is not confined to the evaluation of a specific transition PEP violation probability. 
We consider the predicted PEP violating atomic transition amplitudes, in the context of the $\theta$ Poincar\'e model, whose derivation is outlined in Sections \ref{transition} and \ref{sec:energydep}. The deformation of the standard transition probabilities depends on the typical energy scales of the involved reactions, conditioned by the value of the ``electric like" components of $\theta_{\mu \nu}$.  This enables a fine tuning of the $\theta$ tensor components. The predicted energy dependence of the spectral shape of the whole complex of relevant transitions is accounted for and tested against the measured X-rays distribution, constraining $\Lambda$, for the first time, far above the Planck scale for $\theta_{0i} \neq 0$.
Within a similar theoretical framework the DAMA/LIBRA limit on the PEP violating atomic transition probability \cite{bernabei2009} was analyzed in Ref.~\cite{Addazi:2018ioz}, and a lower limit on the non-commutativity scale $\Lambda$ was inferred as strong as $\Lambda > 5 \cdot 10^{16}$ GeV, corresponding to $\Lambda > 4 \cdot 10^{-3}$ Planck scales.

PEP violating nuclear transitions were also investigated e.g. in  Refs.~\cite{bernabei2009,bellini2010,suzuki1993}. The strongest bound ($\delta^2 < 7.4 \cdot 10^{-60}$) was obtained in Ref.~\cite{bellini2010} searching for PEP violating transitions of nucleons from the 1P$_{3/2}$ shell to the filled 1S$_{3/2}$ shell.
Based on a parametrization of the PEP-violation probability, in terms of inverse powers of the non-commutativity scale, the impact of these experimental results for Planck scale deformed symmetries was investigated in Ref.~\cite{Addazi:2017bbg}. As a result a class of $\kappa$-Poincar\'e and $\theta$-Poincar\'e models was excluded in the hadronic sector. 
Nonetheless, within the context of NCQG models, tests of PEP in the hadronic and leptonic sectors need to be considered as independent. There is no a priori argument why the standard model fields should propagate in the non-commutative space-time background with the same coupling. 
As an example non-commutativity emerges in string theory as a by-product of the constant expectation value of the B-field components, which in turn are coupled to strings' world-sheets with magnitudes that are not fixed a priori.\\

Underground X-ray surveys, searching for atomic transitions prohibited by the PEP, proved in recent years to provide very strong and promising tests of Spin-Statistics \cite{vipepjc2018,piscicchiaentropy2020,Piscicchia:2020kze} and, hence, tests of Quantum Gravity models at unexpectedly high non-commutativity scales. The results of the dedicated VIP-2 Lead experiment, searching for signals of  K$_\alpha$ and K$_\beta$ PEP violating transitions in an ultra radio-pure Roman lead target, are analysed in the framework of NCQG, and summarized in Ref. \cite{PRL_version}. 
We show that the experiment sets critical bounds on
$\theta$-Poincar\'e. In particular, VIP-2 Lead
rules-out $\theta$-Poincar\'e up to $2.6 \cdot 10^2$ Planck scales, if the
``electric like" components of the $\theta_{\mu \nu}$ vector are non-null; $\theta$-Poincar\'e is excluded up to $6.9 \cdot 10^{-2}$ Planck scales if they vanish.
VIP-2 Lead provides the strongest bound, from atomic levels transitions, on this special type of space-time non-commutativity.

\section{Theoretical framework}
\noindent
At a formal level, the discussion on the fate of the spin-statistics theorem is strictly related to the deformation of the Poincar\'e symmetries, which is in turn induced by space-time non-commutativity. Few relevant caveats deserve to be represented with further detail than the one we deployed in the previous section, as these may affect the outputs of the theoretical investigation. We comment here that the arguments advocated in order to prevent the breakdown/deformation of the spin-statistics theorem, as a byproduct of space-time non-commutativity, yield as a consequence that the theory can not be any longer falsifiable.\\ 
This issue has been tackled inspecting the validity of micro-causality, within a quantum framework that is still required to be unitary. Intuitively, one expects the light-cone not to retain any status in a non-commutative theory. It might then be surprising that in the case of the Moyal-plane non-commutativity, which we have referred to in the text as $\theta$-Poincar\'e,  for a certain class of observables, for instance the one involving scalar fields with no time derivative, micro-causality can still hold, namely 
\begin{equation}
[\mathcal{O}(x), \mathcal{O}(y)]_\star  = [:\phi(x)\star \phi(x):, :\phi(y)\star \phi(y):]_\star=0 \,, 
\end{equation}
the equality being supposed to hold for space-like separations, with $: \cdot :$ denoting a fixed ordering of the non-commutative coordinates, and the $\star$-product being defined by 
\begin{equation}
:f(x)\star g(x):= \lim_{y\rightarrow x}\,
f(x)\, e^{\frac{\imath}{2}\, \partial_\mu^x \theta^{\mu\nu} \partial_\nu^y } \, g(y)\,.
\end{equation}

Problems related to a-causality for non-commutative theory were outlined in Ref.~\cite{Seiberg:2000gc}, in which the authors derived the commutator $[\mathcal{O}(x), \mathcal{O}(y)]_\star$. Similarly, using perturbation theory and different definitions of time-ordering, it was pointed out in Ref.~\cite{Bozkaya:2002at} that micro-causality and unitarity cannot simultaneously co-exist in non-commutative field theory. A similar result was obtained in Ref.~\cite{AlvarezGaume:2001ka}, where it was found that SO$(3,1)$ micro-causality is violated, but that SO$(1,1)$ micro-causality in the light-cone still holds when perturbative unitarity does. In Ref.~\cite{Gomis:2000zz} it was then pointed out that non-commutativity among space and time coordinates violates unitarity, unless light-like non-commutativity is considered \cite{Aharony:2000gz}. This problem was supposed to arise because of neglecting massive string states.\\

A different take on the same problem was developed in Ref.~\cite{Chaichian:2002vw}, where it was argued that for pure space non-commutativity micro-causality is preserved. As a by-product of this analysis, Pauli spin-statistics relation was claimed to remain valid in non-commutative quantum field theories. Nonetheless, as pointed in Ref.~\cite{Greenberg:2005jq}, the special case inspected in Ref.~\cite{Chaichian:2002vw} that involves scalar fields with no time derivatives, only make sense in the light-wedge $x_0^2-x_3^2\leq 0$, as already shown in Ref.~\cite{AlvarezGaume:2001ka}. The implications of Ref.~\cite{Chaichian:2002vw} for the spin-statistic theorem contradict the results found in \cite{Greenberg:2005jq}, namely that the equal time 
commutator in \emph{D} space-time dimensions 
\begin{eqnarray}
&&[\mathcal{O}(x), \mathcal{O}(y)]_\star = (e^{-\imath p\cdot x -\imath p'\cdot y }+ e^{-\imath p'\cdot x -\imath p\cdot y } ) \times \nonumber\\
&&\frac{8\imath}{(2\pi)^{2D-1}} \int \frac{d^{D-1}k}{2 E_k}\, \sin \left(-\frac{1}{2} \theta^{ij} \left( k_i(p+p')_j + p_i {p_j}' \right) \right) \times \nonumber\\ 
&& e^{\imath \vec{k}\cdot (\vec{x}-\vec{y}) } \, \cos(\frac{1}{2} \theta^{ij}k_i p_j)\, \cos(\frac{1}{2} \theta^{ij}k_i {p_j}')
\end{eqnarray}
does not vanish at space-like separation. This latter result is instead consistent with previous ones derived in the literature -- see e.g.
Refs.~\cite{Seiberg:2000gc,Bozkaya:2002at,AlvarezGaume:2001ka,Gomis:2000zz,Aharony:2000gz}, and thus hinges toward the confirmation that the spin-statistic theorem is violated.\\


The existence of eventual isomorphism has also been considered in order to instantiate a pure equivalence among commutative and non-commutative quantum field theories \cite{Fiore:2007vg}. Notice that a complete equivalence among the commutative and non-commutative theories, both at the level of the algebra and the co-algebra, would turn non-commutative theory to be un-falsifiable and un-predictive. Consequently, we do not deem as worthy of any phenomenological interest this theoretical approach, since it fails from distinguishing itself from the standard theory and its predictions, and thus decide not to speculate on the implications of this possibility.

\section{Transition amplitudes}\label{transition}
\noindent 
Non-commutative deformation as introduced by the $\star$-product induces deformation in standard processes of QED. The interaction vertex can be extracted from the action
\begin{eqnarray} \label{iv}
    S=  \int d^4x\, \left( -\frac{1}{4} F_{\mu \nu} \star F^{\mu \nu} +
    \bar{\psi} \star (\imath \, \slash \!\!\!\!D \star -m )\psi(x)
    \right)\,,
\end{eqnarray}
with covariant derivative expressed as $D_\mu^\star=\partial_\mu +\imath e A_\mu \star $, $e$ denoting the electron charge, field strength of the electromagnetic field $A_\mu$ provided by $F_{\mu \nu}=\partial_\mu A_\nu -\partial_\nu A_\mu  + \imath e [A_\mu \!\!\! \phantom{a}^\star \!\!, A_\nu]$, with $[A_\mu \!\!\! \phantom{a}^\star \!\!, A_\nu]\equiv A_\mu \star A_\nu- A_\nu \star A_\mu $  and fields undergoing the usual plane-waves expansion
\begin{eqnarray}
    \psi(x)&=  \sum_{s,k} \left[a^{(s)}(k) u^{(s)}(k) e^{-\imath k \cdot x}+b^{\dagger(s)}(k) v^{(s)}(k) e^{\imath k \cdot x}  \right], \nonumber \\
    \bar{\psi}(x)&=  \sum_{s,k} \left[b^{(s)}(k) \bar{v}^{(s)}(k) e^{-\imath k \cdot x}+a^{\dagger(s)}(k) \bar{u}^{(s)}(k) e^{\imath k \cdot x}  \right], \nonumber \\
    \slash \!\!\!\!A(x)&=  \sum_{r,k} \left[ \alpha^{(r)}(k) \slash \!\!\! \epsilon^{(r)}(k) e^{-\imath k \cdot x}+\alpha^{\dagger(r)}(k) \slash \!\!\! \epsilon^{(r)}(k) e^{\imath k \cdot x}  \right], \nonumber 
\end{eqnarray}
with spinor indices $s=1,2$ labelling the set of four independent spinor solutions to the Dirac equation $\{u(k)^{(s)}\,, v(k)^{(s)}\}$, polarization indices $r=1,2$ labelling transverse traceless solution of the Maxwell equations $\epsilon_\mu^{(r)}$ and slash denoting contraction of vectors with $\gamma_mu$ matrices; summation is over discrete momenta $k=(k^{(0)}, \vec{k})$ in some regularization volume. 

The $\star$-product induces deformed commutation and anti-commutation rules:
\begin{eqnarray}
    a^{(s_1)}(p_1)a^{(s_2)}(p_2)=& - e^{\imath p_1 \wedge_\theta p_2 }a^{(s_2)}(p_2)a^{(s_1)}(p_1)\,, \nonumber\\
     b^{(s_1)}(p_1)b^{(s_2)}(p_2)=& - e^{\imath p_1 \wedge_\theta p_2 }b^{(s_2)}(p_2)b^{(s_1)}(p_1)\,, \nonumber\\
      \alpha^{(s_1)}(p_1)\alpha^{(s_2)}(p_2)=& + e^{\imath p_1 \wedge_\theta p_2 }\alpha^{(s_2)}(p_2)\alpha^{(s_1)}(p_1)\,, \nonumber
\end{eqnarray}
 having introduced the antisymmetric product $a\wedge_{\theta} b\equiv a_\mu \theta^{\mu \nu} b_\nu$, and 
\begin{eqnarray}
   a^{(s_1)}(p_1)a^{\dagger (s_2)}(p_2)=& - e^{\imath p_1 \wedge_\theta p_2 }a^{\dagger (s_2)}(p_2)a^{(s_1)}(p_1)  \nonumber \\
   & + 
   \delta_{\vec{p}_1,\vec{p}_2 }\,, \nonumber \\
    b^{(s_1)}(p_1)b^{\dagger (s_2)}(p_2)=& - e^{\imath p_1 \wedge_\theta p_2 }b^{\dagger (s_2)}(p_2)b^{(s_1)}(p_1)\nonumber \\
   & + 
   \delta_{\vec{p}_1,\vec{p}_2 }\,, \nonumber \\
     \alpha^{(s_1)}(p_1)\alpha^{\dagger (s_2)}(p_2)=& + e^{\imath p_1 \wedge_\theta p_2 }\alpha^{\dagger (s_2)}(p_2)\alpha^{(s_1)}(p_1)\nonumber \\
   & 
   + 
   \delta_{\vec{p}_1,\vec{p}_2 }\,. \nonumber 
\end{eqnarray}
Consistently with these rules, asymptotic in-coming and out-going two-fermion states are represented as
\begin{eqnarray}\label{asi}
|p_1,s_1;p_2,s_2\rangle &=& a^{\dagger(s_1)} (p_1) a^{\dagger(s_2)} (p_2)|0\rangle \nonumber \\
&=& e^{\imath p_1\wedge_\theta p_2}  c^{\dagger(s_1)} (p_1) c^{\dagger(s_2)} (p_2)|0\rangle \,,
\end{eqnarray}
and 
\begin{eqnarray} \label{asf}
|p_1',s_1';p_2',s_2'\rangle &=& a^{\dagger(s_1')} (p_1') a^{\dagger(s_2')} (p_2')|0\rangle \nonumber\\
&=& e^{\imath p_1'\wedge_\theta p_2'} c^{\dagger(s_1')} (p_1') c^{\dagger(s_2')} (p_2')|0\rangle \,,
\end{eqnarray}
having introduced ladder operators $c$ and $d$ phase-shifted with respect to $a$ and $b$ by
\begin{eqnarray}
a_{\vec p}= e^{- \frac{1}{2} \imath p\wedge_\theta P} c_{\vec p}\,, \qquad  
b_{\vec p}= e^{- \frac{1}{2} \imath p\wedge_\theta P} d_{\vec p} \nonumber 
\\
{\rm with} \qquad P_\mu = \sum_{\vec p} \left( c^\dagger_{\vec p} c_{\vec p} + d^\dagger_{\vec p} d_{\vec p}
\right) \nonumber 
\end{eqnarray}
and so forth for the hermitian conjugates.\\

It is useful to consider the deformation of the Feynman amplitude $\mathcal{M}$ of the electron-muon scattering process, as recovered from the interaction vertex derived from Eq.~\eqref{iv}and from the asymptotic states in Eqs.~\eqref{asi}-\eqref{asf}. This reads 
\begin{eqnarray}
 \mathcal{M}_\theta &=&    e^{ \frac{\imath}{2} 
 \left( 
 p_1\wedge_\theta p_2- p_1'\wedge_\theta p_2'
 \right)}
\mathcal{M}_0 e^{- \frac{\imath}{2} 
 \left( 
 p_1'\wedge_\theta p_1 + p_2'\wedge_\theta p_2
 \right)}\,,\nonumber \\
  \mathcal{M}_0&=&\frac{e^2}{2 q^2} \left(
 \bar{u}^{s_1'}(p_1')\gamma^\mu u^{(s_1)}(p_1)
  \bar{v}^{s_2'}(p_2')\gamma_\mu v^{(s_2)}(p_2)
 \right) \,,\nonumber
\end{eqnarray}
with transferred momentum $q=p_1'-p_1=p_2'-p_2$. Averaging over the initial spins and summing over final ones, the deformation rearranges, in the cross section $\sigma$, into a multiplicative factor 
\begin{eqnarray}
\sigma_\theta  &=& |\Phi_\theta|^2\sigma_0 \,, \nonumber\\
\Phi_\theta&=&e^{\imath 
( p_1\wedge_\theta p_2 - p_1' \wedge_\theta p_2' )} e^{- \frac{\imath}{2} 
( p_1'\wedge_\theta p_1 - p_2' \wedge_\theta p_2 )} \nonumber\,.
\end{eqnarray}
The first order expansion in the $\theta$ tensor of $|\Phi_\theta|^2$ provides the PEP-violating phase $\phi_{\rm PEPV}=\delta^2$.

\section{Energy dependence of the PEP violation probability in NCQG models}\label{sec:energydep}
\noindent 
Within the framework of NCQG models the space-time non-commutativity is dual to a deformation of the Lorentz/Poincar\'e algebra and, hence, to new space-time symmetry algebras' structures (see e.g. Refs. \cite{6,6b}). Specifically, in order to encode deformations of the Lie algebras of space-time symmetries, the concept of bi-algebras, and the Hopf algebras, must be introduced. These two mathematical concepts, developed from an algebraic abstract viewpoint, have deep physical consequences, since the very same structure of the statistics of fermions and bosons is there encoded. This property sets a direct connection between the microscopic structure of space-time, and the deformation of the spin-statistics relation, more specifically the violation of PEP, which turns to be the observable for NCQG models. More in detail, the $\theta$-Poincar\'e model leads to the  prediction (see Refs.~\cite{Addazi:2017bbg,11,12}) that PEP violations are induced with a suppression $\delta^2 =({E}/\Lambda)^{2}$, where ${E}\equiv {E}(E_1, E_2, \dots)$ is a combination of the characteristic energy scales of the transition processes under scrutiny (masses of the particles involved, their energies, the transitions energies etc.).

For a generic NCQG model deviations from the PEP in the commutation/anti-commutation relations can be parametrized \cite{Addazi:2017bbg} as 
\begin{equation}\label{aadagger}
a_{i}a_{j}^{\dagger}-q(E)a_{j}^{\dagger}a_{i}=\delta_{ij}\, .
\end{equation}
For $\theta$-Poincar\'e models\footnote{In principle, there may exist other models which may have an energy dependence from the electron momentum or mass. It is also interesting to remark that, in general, an angular dependence of the PEPV emission is expected.}, when two electrons of momenta $p_i^\mu=(E_i, \vec{p}_i )$ (with $i=1,2$) are considered, we may introduce the phase $\phi_{\rm PEPV}$ to parametrize the deformation of the standard transition probability $W_0$ into $W_\theta=W_0 \cdot \phi_{\rm PEPV}$. Making explicit $\Lambda$ in the $\theta$-tensor through the relation $\theta_{\mu \nu}= \tilde{\theta}_{\mu \nu}/\Lambda^2$, with $\tilde{\theta}_{\mu \nu}$ dimensionless, the energy scale dependence may be introduced: i) either according to
\begin{equation} \label{alzu}
\phi_{\rm PEPV} = \delta^ 2\simeq \frac{D}{2} \frac{E_N}{\Lambda} \frac{\Delta E}{\Lambda}\,,
\end{equation}
where $D=p_1^0 \tilde{\theta}_{0j} p_2^j + p_2^0 \tilde{\theta}_{0j} p_1^j$, the quantity $E_N\simeq m_N\simeq A \, m_{p}$ denotes nuclear energy and $\Delta E=E_2-E_1$ accounts for the atomic transition energy; ii) or as an energy levels difference, encoding the PEP violating X-ray line energy, namely as in 
\begin{equation} \label{cadu}
\phi_{\rm PEPV} = \delta^ 2 \simeq   \frac{C}{2} \frac{\bar{E}_1}{\Lambda} \frac{\bar{E}_2}{\Lambda}\,,
\end{equation}
where $\bar{E}_{1,2}$ are the energy levels occupied by the initial and the final electrons and $C= p_1^i \tilde{\theta}_{ij} p_2^j$. The former case, discussed in Eq.~\eqref{alzu}, amounts to encode non-commutativity among space and time coordinates, namely $\theta_{0i}\neq 0$, while the latter case, in Eq.~\eqref{cadu}, corresponds to selecting $\theta_{0i}=0$, which ensures unitarity of the $\theta$-Poincar\'e models \cite{AlvarezGaume:2001ka,Gomis:2000zz}. In both cases the factors $D/2$ and $C/2$ can be approximated to unity.     

We further notice that Eq. (\ref{aadagger}) resembles the quon algebra (see e.g. Refs.~\cite{Greenberg:1987aa,greenberg1991particles}), 
in the case of quon fields however the $q$ factor does not show any energy dependence, and it is not related to any quantum gravity model. Thus the q-model necessarily requires a hyper-fine tuning of the $q$ parameter. $q(E)$ is related to the PEP violation probability by 

\begin{equation}
q(E)=-1+2\delta^{2}(E).
\end{equation}

An experimental bound on the probability that PEP may be violated in atomic transition processes, straightforwardly translates into a constrain to the new physics scale $\Lambda$, consistently with the choice of the $\theta_{0i}$ components.

\section{The VIP-2 lead experiment}\label{vip2}
\noindent 
The VIP collaboration is performing high precision tests of the Pauli Exclusion Principle (PEP) for electrons, in the extremely low cosmic background environment of the Gran Sasso underground National Laboratory (LNGS) of INFN.
LNGS provides the ideal conditions for performing  high-sensitivity searches of extremely
low-rate physical processes. The overburden, corresponding to a minimum thickness of 3100 m w.e. (metres water equivalent), yields a reduction of the cosmic muon flux of about six orders of magnitude. As a consequence the main background source, apart from residual cosmic rays, is represented by the $\gamma$-radiation produced by long-lived $\gamma$-emitting primordial isotopes and their decay products, from the rocks of the Gran Sasso mountain and the concrete of the laboratory structure.

The VIP-2 lead experimental apparatus is based on a high purity co-axial p-type germanium detector (HPGe), with a diameter of 8.0 cm and a length of 8.0 cm, surrounded by an inactive layer of lithium-doped germanium of 0.075 mm. The active germanium volume of the detector is 375 cm$^3$. The detector is surrounded by a complex of passive and active shieldings (the latter serves as target material). The passive shielding is composed of an outer pure lead layer (30 cm from the bottom and 25 cm from the sides) and an inner electrolytic copper layer (5 cm). The volume of the sample chamber is of about 15 l ((250 × 250 × 240) mm$^3$). A 1 mm thick air-tight steel housing encloses the shielding structure and the cryostat and is flushed with boil-off nitrogen, from a liquid nitrogen storage tank, to minimize the radon contamination. With the aim to reduce the neutron flux impinging the detector this is also surrounded, on the bottom and the sides, by 
5 cm thick borated polyethylene plates.
The target consists of three cylindrical sections of radio-pure Roman lead, fully surrounding the detector. 
The thickness of the target is about 5 cm, for a total volume $V \sim 2.17\times 10^3\;  \mathrm{cm}^3$. The inner part of the setup is shown in Fig. 1 of Ref. \cite{Piscicchia:2020kze}. A more detailed description of the external 
passive shielding and of the cryogenic and vacuum systems is given in Refs. \cite{neder,heusser2006low}.   
The data acquisition system is a Lynx digital signal analyser controlled via GENIE 2000 personal computer software, both from Canberra-Mirion.

The aim of the measurement is to search for the X-rays signature of PEP-violating K$_\alpha$ and K$_\beta$ transitions in Pb, when the $1s$ level is already occupied by two electrons.
These transitions are shifted, with respect to the standard ones, as a consequence of the shielding effect of the additional electron in the ground state; hence they are distinguishable in precision spectroscopic measurements. 
Let us notice that the deformation of the algebra preserves, at the first order, the standard atomic transition probabilities, the violating transition probabilities being dumped by factors $\delta^2(E)$, hence transitions to the $1s$ level from levels higher then $4p$ will not be considered (see e.g. Ref. \cite{krause} for a comparison of the atomic transitions intensities in Pb). 
The energies of the standard $\mathrm{K}_{\alpha}$ and $\mathrm{K}_{\beta}$ transitions in Pb are given in Table \ref{lines}, where those expected for the corresponding PEP violating ones are also reported. The PEP violating K lines energies are obtained based on a multi configuration Dirac-Fock and General Matrix Elements numerical code \cite{indelicato}, see also Ref. \cite{Elliott:2011cx} where the $\mathrm{K}_{\alpha}$ lines are obtained with a similar technique.


\begin{table}[!h]
\caption{Calculated PEP-violating K${}_{\alpha}$ and K${}_{\beta}$ atomic transition energies in Pb
(column labeled forb.). As a reference, the allowed transition energies are also quoted (allow.). Energies are in keV.}
\label{lines}
\begin{center}
  \renewcommand\arraystretch{1.3}
\begin{tabular}{|c|c|c|}
\hline
\hline
  \textbf{Transitions in Pb}       & \textbf{allow. (keV)}  & \textbf{forb. (keV)}
\\ \hline
   1s - 2p${}_{3/2}$ K${}_{\alpha1}$  &  74.961 & 73.713 \\ \hline
   1s - 2p${}_{1/2}$ K${}_{\alpha2}$  &  72.798 & 71.652 \\ \hline
     1s - 3p${}_{3/2}$ K${}_{\beta1}$  &  84.939 & 83.856  \\ \hline
   1s - 4p${}_{1/2(3/2)}$ K${}_{\beta2}$  &  87.320 & 86.418   \\ \hline
     1s - 3p${}_{1/2}$ K${}_{\beta3}$  &  84.450 & 83.385  \\ \hline

\end{tabular}
\end{center}
\end{table}

The efficiency for the detection of X-rays emitted inside the Pb target, in the energy region of interest (defined in Section \ref{analysis}), was determined by means of Monte Carlo (MC) simulations. To this end the detector was characterised and all of its components have been set into a MC code (Ref. \cite{Boswell:2010mr}) based on the GEANT4 software library (Ref. \cite{Agostinelli:2002hh}).

We present in Section \ref{analysis} the results of a Bayesian analysis, aimed to extract the probability distribution function ({\it pdf}) of the expected number of photons emitted in PEP violating K$_\alpha$ and K$_\beta$ transitions. The analysed data set 
corresponds to a total acquisition time $\Delta t\approx 6.1 \cdot 10^6 \mathrm{s} \approx 70 ~\mathrm{d}$, i.e. about twice the statistics used in Ref. \cite{Piscicchia:2020kze}.   \\

\section{Data Analysis and results}\label{analysis}
\begin{figure}[h]
\centering
\includegraphics[width=\columnwidth]{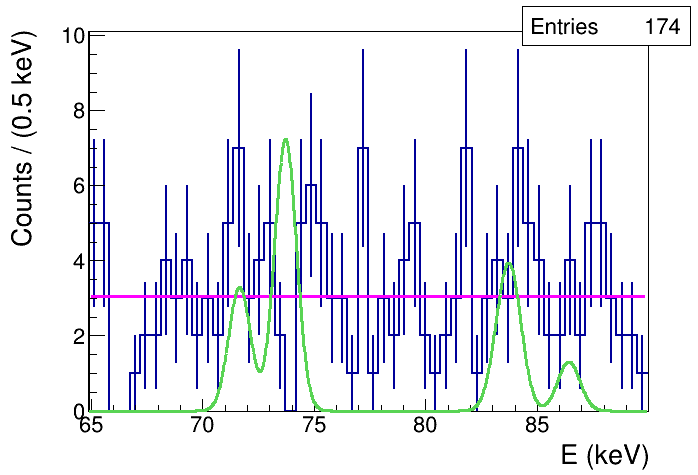}
\caption{
The measured X-ray spectrum, in the region of the K${}_{\alpha}$ and K${}_{\beta}$ standard and violating transitions in Pb, is shown in blue; the magenta line represents the fit of the background distribution. The green line corresponds to the shape of the expected signal distribution (with arbitrary normalization) for $\theta_{0i}\neq 0$.}
\label{spectrum}
\end{figure}

\noindent 
The measured energy spectrum is shown in blue in Figure \ref{spectrum}, in the energy range $\Delta E = (65-90)$ keV. Considered the energy resolution of the detector in this region, which is $\sigma \simeq 0.5$ keV, the interval $\Delta E$ contains the $\mathrm{K}_{\alpha}$ and $\mathrm{K}_{\beta}$ - standard and PEP violating - transitions in Pb. From a detailed analysis of the materials of the setup, these are the only emission lines expected in the selected range $\Delta E$. However, due to the extreme radio-purity of the target even the standard transitions of Pb cannot be recognised over the flat background, which corresponds to an average of about $3$ counts/bin. We also notice that the target contributes to suppress the residual background eventually surviving the external passive shielding complex. 

The aim of the analysis is to extract the upper limit $\bar{S}$ of the expected number of total signal counts, generated by PEP violating K$_\alpha$ and K$_\beta$ transitions in the target. Comparison of $\bar{S}$ with the theoretically expected photons emission, due to PEP violating atomic transitions, provides a limit on the scale $\Lambda$ of the model. 
The conditional \textit{pdf} of the expected number of total signal counts $S$ given the measured distribution - called \textit{data} - is obtained as follows:
\begin{equation}\label{pdf}
P(S | data ) = \int_0^\infty \int_{\mathcal{D}_\mathbf{p}} P(S,B | data, \mathbf{p}) \ d^m\mathbf{p} \ dB
\end{equation}
where the joint posterior distribution of $S$ and the expected number of total background counts $B$ is given by the Bayes theorem 

\begin{equation}\nonumber
    P(S,B | data, \mathbf{p}) =
    \end{equation}
\begin{equation}
     =
     \frac{P(data | S,B,\mathbf{p}) \cdot f(\mathbf{p}) \cdot P_0 (S) \cdot P_0 (B) }{\int P(data | S,B,\mathbf{p}) \cdot f(\mathbf{p}) \cdot P_0 (S) \cdot P_0 (B) \ d^m\mathbf{p} \ dS \ dB} .
    \label{f2}
\end{equation}
In order to account for the uncertainties in the experimental parameters $\mathbf{p}$, which characterize the measurement and the data analysis, an average likelihood is considered, which is weighted with the joint \textit{pdf} of $\mathbf{p}$. $\mathcal{D}_\mathbf{p}$ represents the domain of the parameters' space.  
The likelihood is parametrized as follows
\begin{equation}
 P(data | S,B,\mathbf{p}) = \prod_{i=1}^{N} \frac{\lambda_i (S,B,\mathbf{p})^{n_i} \cdot e^{-\lambda_i (S,B,\mathbf{p})}}{n_i !} \
\end{equation}
where $n_i$ are the measured bin contents. The number of events in the $i$-th bin fluctuates, according to a Poissonian distribution, around the mean value:

\begin{eqnarray}
    \lambda_i (S,B,\mathbf{p}) &=&  B \cdot \int_{\Delta E_i} f_B (E,\bm{\alpha}) \ dE + \\ \nonumber &+& S \cdot \int_{\Delta E_i} f_S (E,\bm{\sigma}) \ dE  
\ 
\end{eqnarray}
$\Delta E_i$ is the energy range corresponding to the $i$-th bin; 
$f_B (E,\bm{\alpha})$ and $f_S (E,\bm{\sigma})$ represent the shapes of the background and signal distributions normalized to unity over $\Delta E$.
Among the experimental uncertainties the only ones which significantly affect $\bar{S}$ are those which characterize the shape of background (parametrized by the vector $\bm{\alpha}$) and the resolutions ($\bm{\sigma}$) at the energies of the violating transitions (the resolutions are reported in Table \ref{sigma}). All the other experimental parameters are affected by relative uncertainties of the order of 1\% (or less), which are neglected, hence $\mathbf{p}=(\bm{\alpha}, \, \bm{\sigma})$.
For the normalized background shape we take a flat distribution, as a result of the fit to the measured distribution described below. 
In order to obtain $f_S (E)$ the expected values of the numbers of PEP violating $K$ transitions, predicted by the model, are to be evaluated.

\begin{table}[!h]
\caption{The table summarizes the resolutions ($\bm{\sigma}$) in keV, at the energies of the PEP violating K${}_{\alpha}$ and K${}_{\beta}$ transitions.}
\label{sigma}
\begin{center}
 \renewcommand\arraystretch{1.3}
\begin{tabular}{|c|c|c|}
\hline
\hline
 \textbf{Transitions in Pb}       & \textbf{$\sigma$ (keV)}  & \textbf{error (keV)} \\ \hline
   K${}_{\alpha1}$  &  0.492 
 & 0.037  \\ \hline
  K${}_{\alpha2}$  &  0.491 & 0.037    \\ \hline
     1s - 3p${}_{3/2}$ K${}_{\beta1}$  &  0.497 & 0.036  \\ \hline
   1s - 4p${}_{1/2(3/2)}$ K${}_{\beta2}$  &  0.498 & 0.036   \\ \hline
     1s - 3p${}_{1/2}$ K${}_{\beta3}$  &  0.497 & 0.036  \\ \hline
\end{tabular}
\end{center}
\end{table}

\subsection{Normalized signal shape}
\noindent 

In order to describe the expected signal shape, corresponding to PEP violating electronic transitions, let us start noticing that electrons belonging to the 2$s$ level are not allowed, at the first order in $\delta^2$, to join the electrons belonging to $1s$. This is motivated by the fact that - at first order - the angular momentum selection rule is preserved by the deformation of the algebra.

We then consider the probability that one electron belonging to the $2p$ shell undergoes the violating K$_{\alpha1}$ transition. This is conditioned to the fact that no other electron performs a transition to the $1s$, otherwise the energy of the violating K$_{\alpha1}$ transition would be shifted, as a result of the further shielding caused by the presence of three electrons in the ground state. The probability is then given by:

\begin{equation}\nonumber
\delta^2(E_{K_{\alpha1}}) \cdot \left[ 1 - 5 \cdot \delta^2(E_{K_{\alpha1}})  \frac{BR_{K_{\alpha1}}}{BR_{K_{\alpha1}} + BR_{K_{\alpha2}}} +  \right.
\end{equation}
\begin{equation}\nonumber
- 5 \cdot \delta^2(E_{K_{\alpha2}})  \frac{BR_{K_{\alpha2}}}{BR_{K_{\alpha1}} + BR_{K_{\alpha2}}} +
\end{equation}
\begin{equation}\nonumber
 -
6 \cdot \delta^2(E_{K_{\beta1}})  \frac{BR_{K_{\beta1}}}{BR_{K_{\beta1}} + BR_{K_{\beta3}}}+
\end{equation}
\begin{equation}\nonumber
\left. -
6 \cdot \delta^2(E_{K_{\beta3}})  \frac{BR_{K_{\beta3}}}{BR_{K_{\beta1}} + BR_{K_{\beta3}}}
-
6 \cdot \delta^2(E_{K_{\beta2}}). \right] = 
\end{equation}
\begin{equation}\label{probka1}
= \delta^2(E_{K_{\alpha1}}) \cdot \left[ 1 +  C_{K_{\alpha1}} \right]. 
\end{equation}
In Eq. \eqref{probka1} the $E_K$ represent the proper energy dependence to be considered in the PEP violation probability, function of the $\theta_{0i}$ choice (see Eqs.~\eqref{alzu}-\eqref{cadu}). The ratios among branching fractions are needed to weight the relative intensities of transitions which occur from levels with the same $(n,l)$ quantum numbers, but different $j$ (e.g. the $2p_{1/2}$ and the $2p_{3/2}$). Such distinction is necessary when the corresponding transition energies can be experimentally disentangled, since for example, transitions as $4p \rightarrow 1s$ can not be resolved (the branching ratios are given in Table \ref{roi}). The term $C_{K_{\alpha1}}$ introduces a second order correction in the searched violating transition probability (and hence on $\Lambda$) which can be neglected. The rate of violating K$_{\alpha1}$ transitions, predicted by the model, from the whole sample of Pb atoms in the target, which would be measured by the detector, is then given by: 
\begin{equation}\label{rate}
\Gamma_{K_{\alpha1}}= \frac{\delta^2(E_{K_{\alpha1}})}{\tau_{K_{\alpha1}}} \cdot \frac{BR_{K_{\alpha1}}}{BR_{K_{\alpha1}} + BR_{K_{\alpha2}}} \cdot 6 \cdot N_{atom} \cdot \epsilon(E_{K_{\alpha1}}). 
\end{equation}
We mean with $\tau_{K_{\alpha1}}$ the lifetime of the PEP-allowed $2p_{3/2}  \rightarrow 1s$ transition (the lifetime for the generic transition will be indicated with $\tau_K$, the lifetimes from Ref. \cite{payne} are summarized in Table \ref{taus}, see also Ref. \cite{reines}). 
The $\epsilon(E_K)$ factors represent the detection efficiencies, for photons emitted inside the Pb target, at the corresponding violating transition energies $E_K$ (the efficiencies are listed in Table \ref{roi}). 
The rate of the generic violating K transition $\Gamma_K$ can be obtained by analogy with Eq. \eqref{rate}.

The probability to observe $n$ violating K$_{\alpha1}$ transitions
in the time $t$ is:
\begin{equation}
    P(n;t) = \frac{(\Gamma_{K_{\alpha1}} \, t)^n \, e^{-\Gamma_{K_{\alpha1}} \, t}}{n!},
\end{equation}
hence the expected number of PEP violating K$_{\alpha1}$ events, predicted by the model, which would be detected in the acquisition time $\Delta t$ is

\begin{equation}\label{expnum}
\mu_{K_{\alpha1}} = \Gamma_{K_{\alpha1}} \cdot \Delta t.
\end{equation}
For a generic violating K transition the expected number of events $\mu_K$ is given by analogy with Eq. \eqref{expnum}.   

Besides the {\it one step} violating K transitions, at the energies which are summarised in Table \ref{lines}, we should also consider two (or more) steps violating transitions populating the same lines. Namely events in which an electron from an atomic shell $i$ undergoes a PEP violating transition to the $n p$ level ($n=2,3,4$), followed by the violating K transition. The  two steps process probability scales as $\delta^2(E_{i \rightarrow np}) \cdot \delta^2(E_{np \rightarrow 1s})$, thus introducing a second order correction to $\mu_K$. The contribution of two (or more) steps violating transitions is then neglected.

Another set of processes to be accounted for consists in subsequent violating transitions from the same atomic shell $n p$ ($n=2,3,4$) to $1s$.
When an electron of the $np$ undergoes a violating transition, it is replaced because of the (non-PEP violating) rearrangement of the atomic shells (the typical transition time scales being much smaller then the $\tau_{K}/\delta^2(E_K)$ lifetime of the violating process). As mentioned above, subsequent violating transitions would populate violating K lines which are shifted in energy with respect to the transitions listed in Table \ref{lines}. The subsequent violating transition probability scales with the product of the $\delta^2$, calculated at the energies corresponding to the two transitions; such processes (as well as higher order ones) are neglected in this analysis. 

$f_S (E)$ is then given by the sum of Gaussian distributions, whose mean values ($E_K$) correspond to the energies of the PEP violating transitions in Pb, and the widths ($\sigma_K$) to the experimental resolutions at the energies $E_K$. The intensities of the violating lines are weighted by the rates $\Gamma_K$ of the corresponding transitions (see Eq. \eqref{rate}): 

\begin{equation}\label{fs}
    f_S (E) = \frac{1}{N} \cdot \sum_{K=1}^{N_K} \Gamma_K \frac{1}{\sqrt{2 \pi \sigma_K^2}}  \cdot e^{-\frac{(E - E_K)^2}{2 \sigma_K^2}}.
\end{equation}

\begin{table}
\caption{The table summarizes the values of the branching ratios of the considered atomic transitions and the detection efficiencies at the energies corresponding to the
K${}_{\alpha}$ and K${}_{\beta}$ forbidden transitions.
}
\label{roi}
\begin{center}
   \renewcommand\arraystretch{1.3}
\begin{tabular}{|c|c|c|}
\hline
\hline
   \textbf{Forb. transitions}  & $BR$ & $\epsilon$
   \\ \hline
    K${}_{\alpha1}$
     & 0.462 $\pm$ 0.009 & $(5.39\pm 0.11)\cdot 10^{-5}$ \\ \hline
    K${}_{\alpha2}$
     & 0.277 $\pm$ 0.006 & $(4.43^{+0.10}_{-0.09})\cdot 10^{-5}$  \\ \hline
       K${}_{\beta1}$
     & 0.1070 $\pm 0.0022$ & $(11.89 \pm 0.24)\cdot 10^{-5}$ \\ \hline
    K${}_{\beta2}$
     & 0.0390 $\pm$ 0.0008 & $(14.05^{+0.29}_{-0.28})\cdot 10^{-5}$  \\
\hline
K${}_{\beta3}$
     & 0.0559 $\pm$ 0.0011 & $(11.51^{+0.24}_{-0.23})\cdot 10^{-5}$  \\
\hline
\end{tabular}
\end{center}
\end{table}
\begin{table}[!h]
\caption{Values of the lifetimes of the PEP-allowed K transitions \cite{payne}.}
\label{taus}
\begin{center}
 \renewcommand\arraystretch{1.3}
\begin{tabular}{|c|c|c|c|c|c|}
\hline
\hline
  $\tau_{K_{\alpha 1}}$  & $1.64 \cdot 10^{-17}$ s \\ \hline
  
  $\tau_{K_{\alpha 2}}$ & $3.6\cdot 10^{-17}$ s
  \\ \hline
  
  $\tau_{K_{\beta 1}}$ & $5.85\cdot 10^{-17}$ s \\ \hline
  
 $\tau_{K_{\beta 2}}$ & $1.42 \cdot 10^{-16}$ s
 \\ \hline
 
 $\tau_{K_{\beta3}}$ & $1.62 \cdot 10^{-16}$ s \\ \hline
\end{tabular}
\end{center}
\end{table}
It is worth noticing that the normalized signal shape depends on the choice of $\theta_{0i}$ (through the proper energy dependence term $E_K$ which is contained in $\delta^2$). Two independent analyses were then performed for the two $\theta_{0i}$ cases, by following the same procedure, in order to set constraints on the $\Lambda$ scale of the corresponding specific model. $f_S$ instead {\it does not} depend on $\Lambda$, since the dependence is re-absorbed by the normalization (see Eq. \eqref{norm}).  
In Eq. \eqref{fs} the sum extends over the number $N_K$ of PEP violating $K$ transitions which are listed in Table \ref{lines}.
The normalization is obtained by the condition:

\begin{equation}\label{norm}
    \int_{\Delta E} f_S (E) dE = 1 \Rightarrow N = \sum_{K=1}^{N_K}
    \Gamma_K.
\end{equation}
As an example, the shape of the expected signal distribution is shown (with arbitrary normalization) as a green line in Figure \ref{spectrum}  for the $\theta_{0i}\neq 0$ choice.

\subsection{Normalized background shape}\label{background}
\noindent 
The normalized background shape is obtained from the best maximum log-likelihood fit to the measured spectrum, excluding 3$\sigma_K$ intervals centered on the mean energies $E_K$ of each violating  transition. The best fit yields a flat background amounting to  $L(E) = \alpha = ( 3.05   \pm 0.29) \ \mathrm{counts/(0.5\,\mathrm{keV})}$, the errors contain both statistical and systematic uncertainties.
The normalized background shape is then:

\begin{equation}
    f_B (E) = \frac{L(E)}{\int_{\Delta E} L(E) \ dE}.
\end{equation}

\subsection{Prior distributions}
\noindent 
For positive values of $B$ we choose a Gaussian prior distribution, with expected value $B_0 = \left<B\right>_G= \int_{\Delta E} L(E) \ dE $ and standard deviation $\sigma_B = \sqrt{B_0}$. Zero probability is assigned to negative values of $B$.
As a check a Poissonian prior was tested for $B$, in this case from the Bayes theorem the expected value is  $\left<B\right>_P=B_0+1$ and $\sigma_B = \sqrt{\left<B\right>_P}$. The upper limit on $\bar{S}$ is found not to be affected by this choice, within the experimental uncertainty.

For what concerns the choice of the prior $P_0 (S)$, considered the a priori ignorance on the value of $S$, we opt for a uniform distribution in the range ($0 \div S_{max}$), where $S_{max}$ represents the maximum value of PEP violating X-ray counts in Pb, compatible with the best independent experimental bound (Ref. \cite{Elliott:2011cx}) on the PEP violation probability.
$S_{max}$ is then obtained from Eq. 3 in Ref. \cite{Elliott:2011cx}, by substituting the number of free electrons in the conduction band of the target, the mean number of interactions and the efficiency with the corresponding parameters which characterise our experimental apparatus (see Tables \ref{roi} and \ref{parameters}).
We obtain $S_{max} \approx 1433$
and the prior on $S$ is

\begin{equation}
    P_0 (S) = 
    \begin{cases}
    \frac{1}{S_{max}}  \qquad 0 \leq S \leq S_{max} \\
    0 \qquad \qquad \text{otherwise}
    \end{cases}
     \ .
\end{equation}
\begin{table}[!h]
\caption{Values of the parameters which characterise the Roman lead target, from left to right: free electron density, volume, mass and number of free electrons in the conduction band.}
\label{parameters}
\begin{center}
 \renewcommand\arraystretch{1.3}
\begin{tabular}{|c|c|c|c|c|c|}
\hline
\hline
  $n_e$(m$^{-3}$)  & $V$(cm$^3$) & $M$(g) & $N_\mathrm{free}$ \\ \hline
  $1.33 \cdot 10^{29}$ & $2.17 \cdot 10^{3}$ & 22300 & $2.89\cdot 10^{26}$ \\\hline
\end{tabular}
\end{center}
\end{table}

\subsection{Lower limits on the non-commutativity scale $\Lambda$}
\noindent

\begin{figure}
    \centering
    \includegraphics[width=\columnwidth]{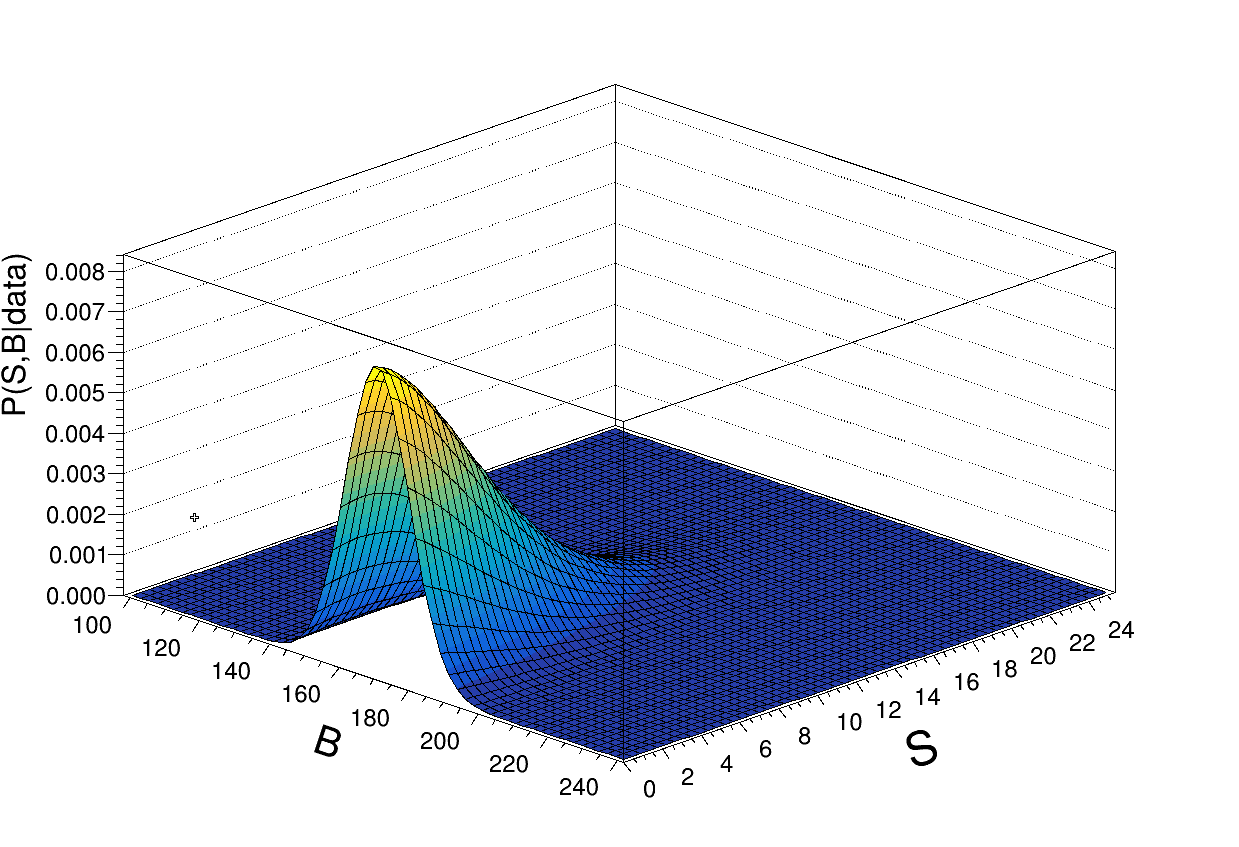}
    \caption{Joint $pdf$ $P(S,B|data)$ of the expected number of total signal and background counts corresponding to $\theta_{0i} \neq 0$.
    .}
    \label{fig:joint}
\end{figure}

The joint $pdf$ $P(S,B|data)$ is shown in Figure \ref{fig:joint} for the case $\theta_{0i} \neq 0$.
The upper limits $\bar{S}$ are calculated, for each choice of $\theta_{0i}$, from the cumulative $pdf$ ($\tilde{P}(\bar{S})$) by solving the integral equation:

\begin{equation}\label{cumulative}
        \tilde{P}(\bar{S}) = \int_0^{\bar{S}} P(S|data) \ dS = \Pi,
\end{equation}
the values obtained for $\bar{S}$, corresponding to a probability $\Pi = 0.9$, are given in Table \ref{limits}, they are affected by a relative numerical error of $\sim 2 \cdot 10^{-5}$.

\begin{table}
\caption{
The table summarizes the upper limits $\bar{S}$ on the expected numbers of signal counts, for a probability $\Pi = 0.9$, and the corresponding lower bounds on the non-commutativity scale $\Lambda$ (in Planck scale unit), for the two choices of $\theta_{0i}$.
}
\label{limits}
\begin{center}
   \renewcommand\arraystretch{1.3}
\begin{tabular}{|c|c|c|}
\hline
\hline
  $\theta_{0i}$   & $\bar{S}$ & \textbf{lower limit on} $\Lambda$ \textbf{(Planck scales)}
   \\ \hline
   
      $\theta_{0i} = 0$
     & 13.2990 & $6.9\cdot 10^{-2}$ \\ \hline
   $\theta_{0i} \neq 0$
     & 18.1515 & $2.6\cdot 10^{2}\,\,\,\,$  \\ \hline
    
\end{tabular}
\end{center}
\end{table}

A dedicated algorithm was developed for the calculation of the posterior $pdf$ in Eq. \eqref{pdf} and of the cumulative distribution function. The numerical integrations are performed by means of Monte Carlo techniques and a detailed description of the numerical tools developed for this analysis is provided in the Appendix.

The comparison of the total expected number of violating transitions $\mu$ predicted by the model and the corresponding upper bound $\bar{S}$, provides a constraint on the lower limit of the scale $\Lambda$, for each choice of $\theta_{0i}$:

\begin{equation}
\mu = \sum_{K=1}^{N_K} \mu_K = 
\frac{\aleph}{\Lambda^2} < \bar{S} \Rightarrow
\end{equation}
\begin{equation}\label{limit}
\Rightarrow \Lambda > \left(  \frac{\aleph}{\bar{S}} \right)^{1/2}.
\end{equation}
The lower limits on $\Lambda$, corresponding to a probability $\Pi = 0.9$, are summarized in Table \ref{limits}. This measurement rules out 
$\theta$-Poincar\'e if the ``electric-like" components of $\theta_{\mu \nu}$ are taken different from zero, since in this case the scale $\Lambda$ of non-commutativity emergence is constrained beyond the Planck scale. If, instead, $\theta_{0i} = 0$ the model is excluded up to $6.9 \cdot 10^{-2}$ Planck scales.

\section{Conclusions}\label{conc}
\noindent 
The results of the analysis of the data collected by the VIP-2 Lead experiment are presented. The measurement was devoted to the search for signals of Pauli Exclusion Principle violating K$_\alpha$ and K$_\beta$ transitions, generated in a high radio-purity Roman lead target. A comparison is performed among the measured spectrum and the expected shape of the violating $K$-lines complex, as predicted in the framework of the $\theta$-Poicaré Non Commutative Quantum Gravity (NCQG) model, based on a Bayesian model. As a result $\theta$-Poicaré is excluded up to 2.6$\cdot 10^2$ Planck scales when the ``electric like" components of the $\theta_{\mu \nu}$ tensor are different from zero, and up to 6.9$\cdot 10^{-2}$ Planck scales if they vanish, thus providing the strongest atomic-transitions based test of the model.

The most interesting phenomenological feature, emerging from the theory, consists on the PEP violating transition amplitude dependence on the characteristic energy scales involved in the investigated reactions (see e.g. Eqs. \eqref{alzu} and \eqref{cadu}). This, in turn, translates in a dependence on the atomic number of the adopted target, thus opening an intriguing scenario demanding a systematic analysis of the data from ongoing   \cite{bernabei2009,abgrall} and forthcoming experiments.
Researches in analogy with what presented in Refs.~\cite{Addazi:2017bbg,Addazi:2018ioz}, addressed on a scan of possible evidences for spin-statistics violation, would supplement the conclusions of this study.
\\    

The VIP collaboration is presently implementing an upgraded experimental setup, based on cutting-edge Ge detectors, aiming to probe $\theta$-Poincaré beyond the Planck scale, independently on the particular choice of the  $\theta_{\mu \nu}$ electric like components. 
%
%
NCQG models, in a large number of their popular implementations, 
are not far to be tested and eventually ruled-out. In this sense, contrary to naive expectations NCQG is not only a theoretical attractive mathematical idea
but also offers a rich phenomenology, 
suitable to be tested in high sensitivity X-ray spectroscopic measurements.\\

\section*{Acknowledgments}
\noindent 
This publication was made possible through the support of Grant 62099 from the John Templeton Foundation. The opinions expressed in this publication are those of the authors and do not necessarily reflect the views of the John Templeton Foundation.
We acknowledge support from the Foundational Questions Institute and Fetzer Franklin Fund, a donor advised fund of Silicon Valley Community Foundation (Grants No. FQXi-RFP-CPW-2008 and FQXi-MGB-2011), and from the H2020 FET TEQ (Grant No. 766900).
We thank: the INFN Institute, for supporting the research presented in this article and, in particular, the Gran Sasso underground laboratory of INFN, INFN-LNGS, and its Director, Ezio Previtali, the LNGS staff, and the Low Radioactivity laboratory for the experimental activities dedicated to the search for spontaneous radiation.
We thank the Austrian Science Foundation (FWF) which supports the VIP2 project with the grants P25529-N20, project P 30635-N36 and W1252-N27 (doctoral college particles and interactions).
K.P. acknowledges support from the Centro Ricerche Enrico Fermi - Museo Storico della Fisica e Centro Studi e Ricerche “Enrico Fermi” (Open Problems in Quantum Mechanics project). 
A.A. work is supported by the Talent Scientific Research Program of College of Physics, Sichuan University, Grant No.1082204112427 \& the Fostering Program in Disciplines Possessing Novel Features for Natural Science of Sichuan University,  Grant No. 2020SCUNL209 \& 1000 Talent program of Sichuan province 2021. 
AM wishes to acknowledge support by the Shanghai Municipality, through the grant No. KBH1512299, by Fudan University, through the grant No. JJH1512105, the Natural Science Foundation of China, through the grant No. 11875113, and by the Department of Physics at Fudan University, through the grant No. IDH1512092/001. 
A.A and A.M. would like to thank Rita Bernabei and Pierluigi Belli for useful discussions on this subject.

\section*{APPENDIX}

\subsection{Details of the integration algorithm}

A dedicated algorithm was developed for solving the following integral equation:

\begin{equation}\nonumber
\tilde{P}(\bar{S}) = \int_0^{\bar{S}} P(S|data) \ dS = 
\end{equation} 
\begin{equation}\label{cumulativa}
 = \int_0^{\bar{S}} \int_0^\infty \int_{\mathcal{D}_\mathbf{p}} P(S,B | data, \mathbf{p}) \ d^m\mathbf{p} \ dB \ dS = \Pi.
\end{equation} 
The integration over $d^m\mathbf{p}$ and $dB$ is performed by using a Monte Carlo method, integration over $dS$ is done by applying the trapezoidal rule.
The range $[0,S_{max}]$ is split in $N_S$ intervals of width $h = S_{max}/N_S$. This defines a vector $\{S_0, S_1, ... , S_{N_S} \}$ such that $S_i = i \cdot h$ with $i = \{0,1,...,N_S\}$. The posterior $pdf$ of $S$ is calculated for each value of $S_i$ by solving the following integrals:

\begin{equation}\nonumber
     P(S_i|data) = \int_0^\infty \int_{\mathcal{D}_\mathbf{p}} P(S_i,B | data, \mathbf{p}) \ d^m\mathbf{p} \ dB = 
\end{equation}
\begin{equation}
  = \frac{P_0(S_i)}{\mathcal{N}} \int_0^\infty \int_{\mathcal{D}_\mathbf{p}} P(data|S_i,B, \mathbf{p}) \cdot f(\mathbf{p}) \cdot P_0 (B) \ d^m\mathbf{p} \ dB,  
     \label{integrale}
\end{equation}
where  $\mathcal{N}$ represents the normalization. The prior of $S$ is defined as:
\begin{equation}
    P_0(S) = \frac{1}{S_{max}} \ \left[ \Theta(S) - \Theta(S - S_{max}) \right]\ ,
\end{equation}
where $\Theta$ is the Heaviside function.
The prior of $B$ is:
\begin{equation}\nonumber
    P_0 (B) = \frac{1}{\sqrt{2 \pi B_0}} \ \exp\left(- \frac{(B-B_0)^2}{2 B_0} \right) \cdot \Theta(B)  = 
\end{equation}
\begin{equation}
   =  \mathcal{G}(B | \ \mathcal{N}_{B_0, B_0}) \cdot \Theta(B) \ ,
\end{equation}
with $B_0 \gg 0$.

The likelihood is weighted over a number $m = N_K + 2$ of experimental parameters $\mathbf{p}=(\bm{\alpha}, \, \bm{\sigma})$. The joint $pdf$ of $\mathbf{p}$ is a multivariate Gaussian:
\begin{equation}
    f(\mathbf{p}) = \mathcal{G}(\mathbf{p} | \ \mathcal{N}_{\mathbf{p}_0, \mathbf{\Sigma}})  
    \ ,
\end{equation}
whose mean values $\mathbf{p}_0=(\bm{\alpha}_0, \, \bm{\sigma}_0)$ are defined by the fit parameters $\bm{\alpha}_0$ 
given in Section \ref{background} and by the resolutions $\bm{\sigma}_0$ of the transition lines reported in Table \ref{sigma}. The $m \times m$ covariance matrix is $\bm{\Sigma} = \mathrm{diag}(\bm{\Sigma_{\alpha}}, \, \bm{\Sigma_K})$, with $\bm{\Sigma_{\alpha}}$ and $\bm{\Sigma_K}$ the corresponding uncertainties. 

\subsection{Monte Carlo integration}
\noindent 
Eq.~(\ref{integrale}) can be rewritten as
\begin{equation}\nonumber
\begin{aligned}
     P(S_i|data) &= \frac{P_0(S_i)}{\mathcal{N}} \int_{-\infty}^{+\infty} \int_{\mathcal{D}_\mathbf{p}} P(data | S_i,B,\mathbf{p}) \cdot \\ &  \cdot  f(\mathbf{p}) \cdot
     \mathcal{G}(B | \ \mathcal{N}_{B_0, B_0}) \cdot \Theta(B) \ dB =\\
     &= \frac{P_0(S_i)}{\mathcal{N}} \ E_{h}[P(data|S_i,B) \ \Theta(B)],
\end{aligned}
\end{equation}
where $h(B,\mathbf{p}) = \mathcal{G}(B | \ \mathcal{N}_{B_0, B_0}) \cdot f(\mathbf{p})$ is chosen as the importance sampling distribution and $E_{h}$ represents the expectation value calculated over $h$. A set $\{(B_1, \mathbf{p}_1),...,(B_{N_h},\mathbf{p}_{N_h})\}$ of $N_h$ values for the stochastic variables $B$ and $\mathbf{p}$ are extracted according to the distribution $h(B,\mathbf{p})$. 
The $N_S + 1$ values of the posterior $pdf$ of $S$ are then estimated as follows:

\begin{equation}\nonumber
    P(S_i|data) = \frac{P_0(S_i)}{\mathcal{N}} \ \frac{1}{N_h} \sum_{j=1}^{N_h} P(data|S_i,B_j,\mathbf{p}_j) \ \Theta(B_j) =
\end{equation}
\begin{equation}\label{IS1}
 = \frac{P_0(S_i)}{\mathcal{N}} \ \frac{1}{N_h} \sum_{j=1}^{N_h} \mathcal{I}(S_i,B_j,\mathbf{p}_j) .
\end{equation}

\subsection{Calculation of the cumulative $\tilde{P}(\bar{S})$}
\noindent 
By applying the trapezoidal rule the normalization integral is approximated by:

\begin{equation}
    \mathcal{N} =  \frac{1}{2 N_S} \frac{1}{N_h} \sum_{i=1}^{N_S} \sum_{j=1}^{N_h} [\mathcal{I}(S_{i},B_j,\mathbf{p}_j) + \mathcal{I}(S_{i-1},B_j,\mathbf{p}_j)] .  \label{normalizzazione} 
\end{equation}
The $i$-th value of the posterior is then obtained as

\begin{equation}
    P(S_i | data) = \frac{ \frac{2 N_S}{S_{max}} \ \sum_{j = 1}^{N_h} \mathcal{I}(S_i,B_j,\mathbf{p}_j)}{\sum_{i=1}^{N_S} \sum_{j=1}^{N_h} [\mathcal{I}(S_{i},B_j,\mathbf{p}_j) + \mathcal{I}(S_{i-1},B_j,\mathbf{p}_j)]},
\end{equation}
from which the cumulative distribution can be sampled as well, for a given $S = S_n$ with $n = \{0,1,...,N_S\}$ it is given by:

\begin{equation}
    \tilde{P}(S_n) = \frac{\sum_{i=1}^{n} \sum_{j=1}^{N_h} [\mathcal{I}(S_{i},B_j,\mathbf{p}_j) + \mathcal{I}(S_{i-1},B_j,\mathbf{p}_j)]}{\sum_{i=1}^{N_S} \sum_{j=1}^{N_h} [\mathcal{I}(S_{i},B_j,\mathbf{p}_j) + \mathcal{I}(S_{i-1},B_j,\mathbf{p}_j)]} .
\end{equation}
To conclude the value of $\bar{S}$ is obtained from a linear interpolation, among the two values  $S_k$ and $S_{k+1}$ for which

\begin{equation}
    \tilde{P}(S_k) < \Pi < \tilde{P}(S_{k+1}).
\end{equation}

\end{document}